\documentclass[conference]{IEEEtran}
\usepackage{graphicx}
\usepackage{algorithm}
\usepackage{algorithmicx}
\usepackage{epstopdf}
\usepackage[noend]{algpseudocode}
\usepackage{cite}
\usepackage{color}
\usepackage{amsmath,amssymb,amsfonts,amsthm}
\usepackage{bm}
\usepackage{enumerate}
\usepackage{multirow}
\usepackage{array, booktabs}

\begin{document}
\title{Ecology-Based DoS Attack in Cognitive Radio Networks}

\author{\IEEEauthorblockN{Shin-Ming Cheng}
\IEEEauthorblockA{Department of Computer Science and Information Engineering\\
National Taiwan University of Sciecne and Technology \\
Taipei, Taiwan\\
Email: smcheng@mail.ntust.edu.tw}
\and
\IEEEauthorblockN{Pin-Yu Chen}
\IEEEauthorblockA{Department of Electrical Engineering and Computer Science\\
University of Michigan \\ Ann Arbor, USA \\
Email: pinyu@umich.edu}}

\maketitle

\begin{abstract}
Cognitive radio technology, which is designed to enhance spectrum utilization, depends on the success of opportunistic access, where secondary users (SUs) exploit spectrum void unoccupied by primary users (PUs) for transmissions. We note that the system behaviors are very similar to the interactions among different species coexisting in an ecosystem. However, SUs of a selfish nature or of misleading information may make concurrent transmissions with PUs for additional incentives, and thus disrupt the entire ecosystem. By exploiting this vulnerability, this paper proposes a novel distributed denial-of-service (DoS) attack where invasive species, i.e., malicious users (MUs), induce originally normal-behaved SUs to execute concurrent transmissions with PUs and thus collapse the cognitive radio network. We adopt stochastic geometry to model the spatial distributions of PUs, SUs, and MUs for the analysis of the mutual interference among them. The access strategy of each SU in the spectrum sharing ecosystem, which evolves with the experienced payoffs and interference, is modeled by an evolutionary game. Based on the evolutionary stable strategy concept, we could efficiently identify the fragile operating region at which normal-behaved SUs are eventually evolved to conduct concurrent transmissions and thus to cause the ruin of the network.
\end{abstract}



\section{Introduction}
\label{sec_intro}

Facing the dynamic and considerable wireless resource demands, the typical spectrum management approach of allocating fixed spectrum bands to licensed users is criticized by its underutilization. \textit{Cognitive radio} (CR) receives a lot of attentions in both academic and industrial areas since it increases spectrum utilization by exploiting local sensing information and agility/dynamic spectrum access. Specifically, unlicensed secondary users (SUs) sense surrounding environment and exploit the spectrum hole unoccupied by licensed primary users (PUs) for secondary transmission with minimal interference to PUs~\cite{Goldsmith09}. In this way, responsibility for avoiding harmful interference is shifted from the regulatory with fixed mandate to equipments that can adapt at runtime~\cite{Atia08}. Such a distributed and light-handed regulation assumes that SUs comply with the sharing etiquette such as evacuating the spectrum upon sensing primary transmission to ensure the normal operation of CR.

In this paper we propose to analyze the stability of a CR network from the perspective of evolution from biology. Using terminologies from ecological biology, the entire CR network can be viewed as an ecosystem~\cite{Mazurczyk16}, where SUs of different spectrum access strategies are different species, the common resource shared by all species are the vacant wireless spectrum, and the fitness of an SU is the utility it received given the profile of spectrum access strategies of all SUs. Notably, the well-known ``survival of the fittest'' phenomenon applies to CR network as well. The autonomous spectrum access behavior of CR leads to the optimal spectrum access strategy that maximizes the utility, which is exactly analog to developing the best fitting rule to  survive in an ecosystem. However, such evolutionary stability may be disrupted or weakened when ``invasive species'' (such as malicious attackers) come into play~\cite{Grosholz02}. SUs controlled by malicious attackers can be viewed as a new species with the objective of disrupting the entire ecosystem. As will be studied in this paper, the presence of such a disruptive species may consume many resources such that the spectrum access strategies of normal SUs become more aggressive due to the adaption rule, and eventually resulting in less evolutionary stability, or even worse, the extinction of SUs with regular spectrum access strategies.

To realize such attack, we investigate the possible target in CR for adversarial or malicious users (MUs)  aiming to launch denial-of-service (DoS) attacks~\cite{Baldini12,Attar12,Fragkiadakis13,Sharma15}. Typically, the critical functionalities for CR ecosystems, including spectrum sensing, agile radio, and light-handed regulation, are the possible candidates since once these functionalities fail, SUs are not able to communicate effectively. For example, MUs can directly jam the victim by injecting interference or deceive SUs into believing that there is a PU by emulating the signal characteristics of the PU, thereby evacuating the occupied spectrum~\cite{Chen06}. Moreover, the \textit{liability} rule is vulnerable to the selfish and greedy users aiming to maximize their own private benefits. Since complying with the rule results in less transmission opportunities, such SUs may not want to invest efforts to follow the rule and thus will transmit simultaneously with PUs~\cite{Faulhaber08}.

In distributed cognitive radio ad hoc networks (abbreviated as CRNs), the situation is more challenging because enforcing the compliance of rules is virtually impossible. With the sensing capability, SUs can acquire more information from the surrounding environment than PUs, which results in \textit{information asymmetry} between them. Without the threat of being easily detected and punished from a central authority, SUs with much more information and agile radio may cheat intentionally. In this case, the well-known self-enforcement among SUs~\cite{Tse07} can not be achieved since the balance between their own quality of service and the interference they are causing to other users is upset by the strong incentives of concurrent transmissions. This leads to a \textit{tragedy of the commons} dilemma~\cite{Hardin68} where a shared limited resource is depleted beyond a recoverable level by individuals pursuing their own best interests.

Inspired by ecological biology~\cite{Vincent05}, PUs, SUs, and MUs with different behaviors could be considered as different species in an ecosystem. In~\cite{Cheng11}, we discover evolutionary dynamics of CRN such that misbehaved SUs evolve to take over the entire networks. Taking advantages of selfish nature, this article further proposes an ecology-based DoS attack~\cite{Mazurczyk15} from adversary's perspectives. In such a distributed DoS attack, the randomly distributed MUs cooperatively induce originally normal-behaved SUs to conduct concurrent transmissions by showing them significant incentives to do that. Consequently, both PUs and SUs suffer heavy interference and entire or a substantial part of the network collapses. From ecological aspect, such an attack can be interpreted as a process that the native species perceive incentives from the behaviors of invasive non-indigenous species, believe their behaviors are more fitting to environment, and thus evolve to become one of them.

We apply stochastic geometry~\cite{Haenggi09} as the framework for modeling the spatial distributions of PUs, SUs, and MUs to quantitatively analyze the mutual interference among them, which affects the quality of concurrent transmissions and influences the behaviors of SUs. Via evolutionary game modeling~\cite{Tembine10,Jiang13} where the access strategy of each SU evolves with the experienced interference and payoff, we can analyze the time dynamics of (mis)behaved SUs and understand the role of MUs in decline of the population of behaved SUs. Based on the evolutionary stable strategy (ESS) concept, we identify the robust operating region where SUs can self-enforce themselves to comply with the sharing rules, which can aid in the widespread deployment of CR technology. The numerical results show that fewer MUs are required by our approach to create more severe damage in comparison to existing direct jamming attacks. With the additional information acquired by sensing, the selfish SUs are easily induced to pursue incentives and cause the ruin of the network.

The remainder of this paper is organized as follows. Section~\ref{sec_back} introduces basic features of CRN, and surveys existing DoS attacks in CRN and emphasizes our contributions and novelties. The proposed ecology-based DoS attack consisting of three phases is described from both engineering and evolutionary aspects in Section~\ref{sec_att}. Section~\ref{sec_ana} applies an evolutionary game to analyze the dynamics of access strategy of SUs during each phase of the attack. Numerical results are provided in Section~\ref{sec_num} and Section~\ref{sec_con} concludes this paper.

\section{Background and Related Works}
\label{sec_back}

\subsection{CR Ecosystems}
\label{sec_ope}


Using terminologies from ecological biology, the entire CR network can be viewed as an ecosystem~\cite{Mazurczyk16}, where SUs of different spectrum access strategies are different species, the common resource shared by all species are the vacant wireless spectrum, and the fitness of an SU is the utility it received given the profile of spectrum access strategies of all SUs. Two essential operations must be ensured to guarantee the success of the opportunistic access of CR ecosystem:`
\begin{itemize}
\item An SU must collect and process information about coexisting users within its spectrum, which requires advanced sensing and signal-processing capabilities.
\item An SU must follow the sharing rule to allocate the resource without or with constrained interfering the PUs.
\end{itemize}


To violate availability of CR ecosystems, the invasive species, i.e., MUs, can easily force these two CR operations nonfunctional. SUs affected by MUs can be viewed as a new species (or mutuants) with the objective of disrupting the entire ecosystem.  Moreover, the interoperability and dynamic nature in CR causes a burden on identity authentication and makes environment more harsh. Thus, MU could easily take on multiple identities and behaves as multiple distinct Sybil users~\cite{Tan11C1}. The reconfiguration capability of CR needs downloading of software module (e.g., waveform or radio application) and thus is vulnerable to the malicious codes~\cite{Baldini12}. The above unique features facilitate the realization of DoS attack and increase the difficulty to catch the MUs.

\subsection{Attack Taxonomy in CRN}
\label{sec_tax}

\begin{table*}[!t]
\caption{DoS attacks in cognitive radio networks} \centering
\begin{tabular}{@{}cccccccc@{}}
\toprule \multicolumn{3}{r}{} & \multicolumn{3}{c}{Target of attack} & \multicolumn{2}{c}{Action of attack}  \\
\cmidrule(r){4-6} \cmidrule(l){7-8} \multicolumn{3}{l}{} & SUs & PUs & Service & Interference & False feedback \\
\midrule
\multirow{4}{*}{Cause of} &  Wireless feature & Jamming  & o & o &  & o & \\
\cline{2-8} & Collaborative sensing & SSDF~\cite{Chen08} & o &  &  &  & o\\
\cline{2-8} & \multirow{2}{*}{Sharing rules} &  PUEA~\cite{Chen06} & o &  & & o & o  \\
\cline{3-8} & & our work & o & o & o & o & \\
\bottomrule
\end{tabular} \label{table_taxo}
\end{table*}

In such harsh environment with possible vulnerabilities introduced by new features of CR, security threats and DoS attacks are receiving lots of attention. Survey of security threats and DoS attacks in CRN are provided in~\cite{Baldini12,Attar12,Fragkiadakis13,Sharma15} using different taxonomies. This paper further classifies the DoS attacks with a novel taxonomy, which identifies the major dimensions of an attack: targets chosen, vulnerabilities exploited and actions taken.
\begin{itemize}
\item Target of attack: This dimension consists of SUs, PUs, and the entire CRN service/technology. Typically, DoS attacks in CRN try to deny the communication for legitimate SUs even when the system resources are available. In some cases, MUs target on PUs by directly jamming or simulating SUs to cause interference. By incurring destructive consequences on the entire network (including both PUs and SUs), an attack could have a negative impact on peoples' acceptance and adoption of CR and thus forestall the widespread deployment of CR technology.
\item Cause of attack: This dimension includes the vulnerabilities in CRN that could be exploited by the adversary, such as shared wireless media, collaborative sensing, learning mechanism, and the light-handed sharing rules.
\item Action of attack: The communication of the victim can be disabled by directly interfering to the receiver from MU. Alternatively, MU could intentionally feedback false information to fool legitimate users for the potential gain.
\end{itemize}

Table~\ref{table_taxo} classifies the existing DoS attacks in CRN according to proposed taxonomy and details are described as follows.

\begin{itemize}
\item Jamming Attack. The shared nature of the wireless medium is vulnerable to jamming attacks. An MU can randomly jam and disrupt any ongoing PU or SU communications by injecting interference. To obtain higher benefits, multiple MUs are suggested to launch such attacks to a target in a coordinated manner. Recent research~\cite{Ray11} adopts stochastic game extended from Markov decision process to defend against jamming attacks.

\item Primary User Emulation Attack. By exploiting the sharing rules that protect PUs, an MU can compel an SU to vacate the occupied spectrum by mimicking the PU~\cite{Chen06}. Then, the MU benefits from the exclusive usage of resources released by the SU and the SU are prohibited from exploiting vacant resources. This type of attack is more efficient than conventional jamming since only low power is needed to dominate the frequency band. To detect and combat PUEA, a multitude of studies have been proposed to identify the masquerading threat through signal analysis~\cite{Xin14,Thanh15}

\item Spectrum Sensing Data Falsification Attack. The accuracy of spectrum sensing is improved by leveraging the observations from multiple SUs. In this case, an MU may send other SUs dishonest reports to lead false conclusion on the presence or absence of PUs. Consequently, the MU benefits from the specific band evacuated by SUs or PUs suffer the harmful interference caused by SUs~\cite{Chen08}. Substantial efforts have been spent to catch the stealthy MU and ignore its reports by using reputation metrics calculation, game theory, abnormality detection, and Bayesian analysis~\cite{Zhang15}.
\end{itemize}

In~\cite{Baldini12,Fragkiadakis13,Attar12,Sharma15}, researchers point out a possible security threat in CRN due to selfish and abnormal behaviors, where MUs or SUs disrupt the sharing rules and access the spectrum without authorization~\cite{Bian13,Jo13}. To enforce compliance on liability rule in CRN,~\cite{Sahai08} develops a model to investigate whether a jail-based punishment is sufficient to convince an SU to respect sharing rules. In the subsequent effort~\cite{Atia08}, a coding architecture is proposed to identify SUs who cause harmful interference. Our previous research~\cite{Cheng11} proposes a model to quantitatively analyze the effects of breaking the liability rule on the ecological survivability of CRN. In contrast to the existing approaches, this attack stimulates SUs to cause interference to both PUs and SUs. The resulting network breakdown may forestall the widespread deployment of CR. Instead of directly injecting interference, MUs induce legitimate SUs to disrupt. As the avalanche effects in the existing cascade-based attack in complex networks~\cite{Motter02}, adversary can cause the same damage in targeted networks by injecting a relatively small amount of traffic. By avoiding a direct attack, the probability that an MU to be detected and identified decreases \cite{CPY14JIOT,CPY_sequential}.

\section{The Proposed Ecology-Based DoS Attack}
\label{sec_att}

This section discusses the proposed attack from both evolutionary and engineering perspectives. In evolutionary aspect, users with different access strategies and behaviors are regarded as different species. Originally, all PUs and most of SUs behave normally, which are considered as native species in a balanced ecosystem~\cite{Cheng11, Mazurczyk16}. The stealthy MUs involving later try to prevent usable reception on victims by forcing their received SINR lower than the threshold. The invasive species, i.e., MUs, simply act as mutants who conduct concurrent transmissions with PUs. The reaction of SUs can be interpreted as a process that the native species perceive misbehavior of mutants (i.e., MUs), believe their behaviors have more fitness to environment, and thus evolve to be one of them.

The primary goal of MUs is to prevent usable reception on victims and thus MUs are assumed to be irrational, that is, MUs do not care at all their quality of service and consuming power. Generally, we assume that MUs use the same transmission device as normal SUs to disguise their real identities. However, the capability of MUs is much more powerful than legitimate SUs. For example, an MU knows the payoff of every user and knows a user is malicious or not~\cite{Baras08}. Moreover, an MU can establish out-of-band fast channels to collude with other MUs as well as to control Sybil identities and its zombies to mount a cooperative attack~\cite{Tan11C1}.

\begin{figure*}[!t]
    \centering
    \includegraphics[width=5in]{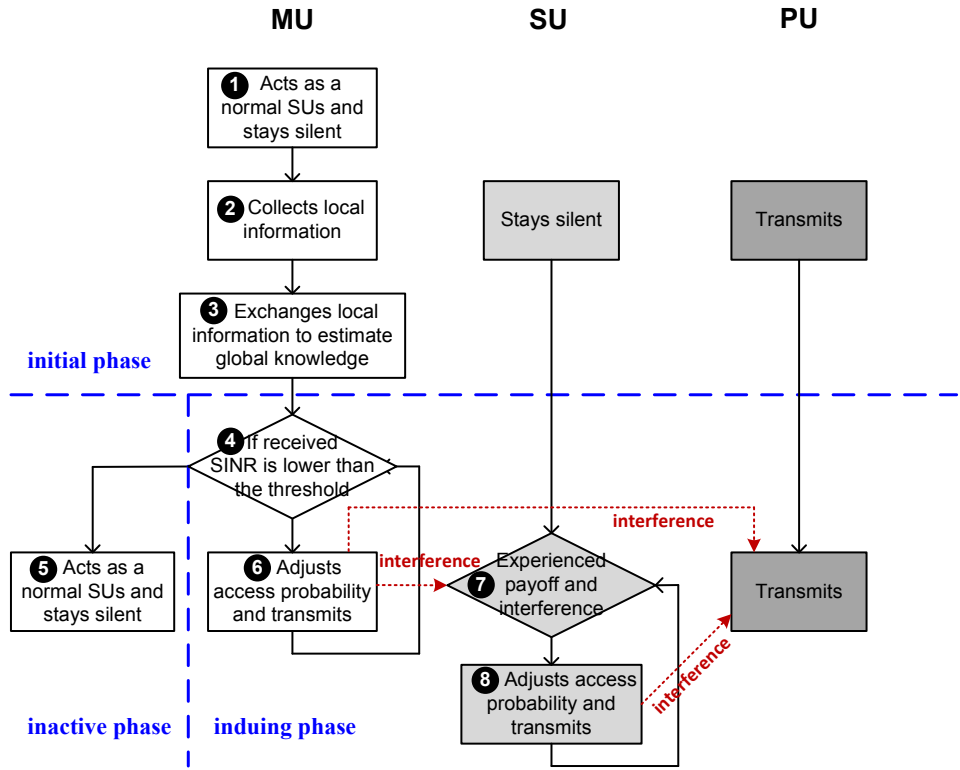}
    \caption{The flow chart of attack procedure for MUs and the corresponding actions in SUs and PUs during each phase of the attack.}
    \label{fig_phase}
\end{figure*}

To disrupt the reception on the primary target, PUs, MUs always perform the attack during the period when PUs are actively transmitting. As illustrated in Figure~\ref{fig_phase}, the attack consists into several phases. In the \textit{initial phase} of the attack, MUs act as normal SUs and stay silent. In the meanwhile, MUs passively collect local environment parameters such as numbers of surrounding PUs and SUs. By exchanging the acquired information, MUs could estimate some global knowledge about networks (such as user density) and accordingly determine whether launching attack is efficient or not (see Steps 1-3 in Figure~\ref{fig_phase}).

Once MUs decide to perform the attack, they enter the \textit{inducing phase} by acting as unintentional malfunctioning SUs and injecting interference to surrounding users. MUs also apply slotted ALOHA as MAC protocol and randomly execute jamming to increase the difficulty to be detected. SUs detect the existence of other SUs (pretended by MUs) and perceive the concurrent transmissions arisen by MUs (see Step 7 in Figure~\ref{fig_phase}). Due to the selfish nature, SUs are beguiled to change their access strategy by increasing their access probability to much larger than $0$ while ignoring PUs might be transmitting (see Step 8 in Figure~\ref{fig_phase}).

MUs with enhanced capability and global information could estimate when SUs have been trapped in the transgression and made primary transmission failed. By estimating the received SINR received from all users, MUs could determine if SUs start transmitting (see Step 4 in Figure~\ref{fig_phase}). If SUs do transmit, MUs transient into the final phase, \textit{inactive phase}. In this phase, MUs could act as regular SUs by deactivating the malicious attempt and choosing the access strategy according to SUs' payoff function. Alternatively, MUs could just stop transmitting to save power and to hide identities (see Step 6 in Figure~\ref{fig_phase}).

The unique features of this novel attack are described as follows.
\begin{itemize}
\item In contrast to the existing approaches, this attack stimulates SUs to cause interference to both PUs and SUs. The resulting vital outbreak may forestall the widespread deployment of CR.

\item In addition to exploiting the sharing rules to mount a attack, the proposed attack utilizes the information asymmetry between SUs and PUs as well as the nature of selfishness to entice SUs to make concurrent transmissions and thus interfere with the targets.

\item Instead of directly injecting interference, MUs induce legitimate SUs to disrupt. As the avalanche effects in the existing cascade-based attack in complex networks~\cite{Motter02}, adversary would be able to cause the same damage in targeted networks by injecting a relatively small amount of traffic. By decreasing the degree of attacking directly, the probability that an MU to be detected and identified is decreased.
\end{itemize}

\section{Analysis of SU Behaviors under the Attack}
\label{sec_ana}

\subsection{Stochastic Geometry}

The performance of communications among spatially scattered nodes in wireless networks is highly constrained by the received power and interference. To model the interference, the spatial distribution and transmission features of the interferers as well as the propagation characteristics of the media shall be addressed. By adopting spatial point process to model the node locations, the interference distributions and link outages can be consequently analyzed. This article applies the homogeneous Poisson point process (PPP) to model the random locations of transmitters and receivers to get tractable analytical result on performance of cognitive radio networks. With PPP, the probability that there are $n$ nodes in $A$ is given by the Poisson distribution (i.e., $(\lambda A)^n e^{–\lambda A}/n!$) where $\lambda$ is the density of nodes in a unit area.

We consider an ad hoc network on a slotted system consisting of PUs, SUs and MUs utilizing the same spectrum. The spatial distributions of primary transmitters (PTs) and SUs are assumed to follow homogeneous Poisson point processes (PPPs) with densities $\lambda_{PT}$ and $\lambda_{SU}$, respectively. Each PT has transmission power $P_{PT}$ and a dedicated primary receiver (PR) located at a fixed distance $r_{PT}$ with an arbitrary direction. The spatial distribution of PRs also forms a PPP with the same density $\lambda_{PT}$ correlated with that of PTs. SUs use fixed transmit powers of $P_{SU}$ and transmission ranges of $r_{SU}$.

The SUs with CR capability are assumed to sense and distinguish signaling from PUs and surrounding SUs perfectly. We consider the interweave paradigm for CR adaptation, that is, SUs should comply with the sharing rule that PUs can not be interfered at all. This implies that SUs only use the spectrum that is not temporarily used by PUs and are obligated to evacuate the spectrum upon sensing primary transmission. The spatial distribution of MUs is also assumed to follow a homogeneous PPP with density $\lambda_{MU}$, MUs use fixed transmit powers of $P_{MU}$ and transmission ranges of $r_{MU}$.

We consider path loss attenuation effects and Rayleigh fading with unit average power in our channel model. The path-loss exponent of transmission is denoted by $\alpha$. Thus the power at receiving side is denoted as $PL(d) = \mathcal{G} d^{-\alpha}$, where $d$ is the distance between the transmitter and receiver and $\mathcal{G}$ is the channel power gain of the desired link which is exponentially distributed with unit mean. Denote the signal-to-interference-plus-noise ratios (SINRs) observed by a PR and an SU as $\gamma_{PR}$ and $\gamma_{SU}$, respectively. The primary and secondary transmissions are respectively successful if $\mathbb{P}[\gamma_{PR} < \eta_{PR}] \le \epsilon_{PR}$ and $\mathbb{P}[\gamma_{SU} < \eta_{SU}] \le \epsilon_{SU}$, where $\eta_{PR}$ and $\eta_{SU}$ are respectively the SINR thresholds of a PR and an SU, and $\epsilon_{PR}$ and $\epsilon_{SU}$ are respectively the outage constraints of a PR and an SU.

\begin{figure}[!t]
    \centering
    \includegraphics[width=3.3in]{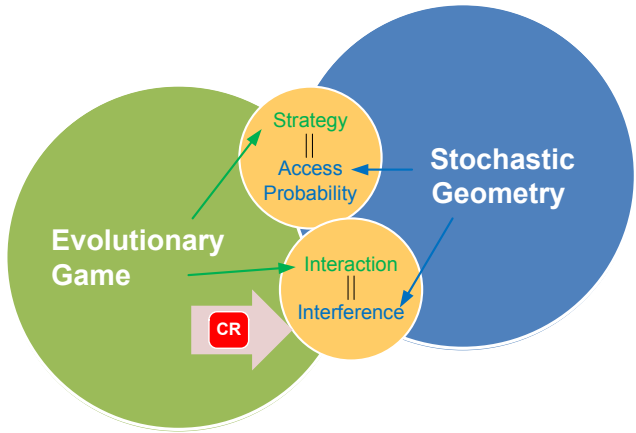}
    \caption{The combination of stochastic geometry and evolutionary game theory for the analysis of ecology-based DoS attack}
    \label{fig_combin}
\end{figure}

\subsection{Evolutionary Game}

Although stochastic geometry provide analytically tractable tools for modeling the tradeoffs between the aggregate interference and spatial contentions, in reality wireless devices are able to adjust the spectrum access parameters according to the experienced spectrum status in order to maximize the system throughput. In the ad hoc environment with one channel and slotted Aloha MAC protocol, access probability control is an instinctive solution for CR adaptation in interweave paradigm. Each SU chooses from the same set of strategies and each strategy corresponds to performing transmission with certain probability. Take two strategies as an example, SUs can take actions of staying silent and transmitting for sure. These settings further complicate the system performance analysis when such time dynamics are involved.

In addition to performing exhaustive experiments to specify these aspects, evolutionary game models have been introduced to investigate the interactions between the overall system performance and the time-evolving access strategies by relating the experienced performance (e.g., SINR, data loss rate) to the game payoff. When the PU is transmitting, the payoff function of an SU depends on the SU's behavior under the greedy nature. If an SU obeys the sharing rule and stays silent when a PU is transmitting, it spends zero cost but obtains some rewards $\kappa$, Furthermore, an SU may not want to break the sharing rule if there is no other accomplice currently conducting transmissions. This implies that payoff for secondary transmission is zero if currently no other SU is perceived. According to \textit{broken windows theory}, the occurrence of other SUs who break the rule encourage its own desire to behave abnormally. In this case, the payoff function depends on if the secondary transmission is successful or not (i.e., if the received SINR at SU is higher than the threshold). We denote $\nu$ as the cost of unsuccessful transmission and $\delta$ as the incentive of successful transmission.

Since cognitive radio networks are deployed in an ad hoc manner, at each stage each CR device tends to adjust its access strategy based on the experienced payoff to maximize its utility, which therefore forms an evolutionary access game. As shown in Figure~\ref{fig_combin}, to address the interactions between the spatial contentions and the evolutionary access strategies, we use stochastic geometry to characterize the aggregate interference according to the spectrum access strategies at each stage. The spectrum access strategies are associated with the parameters of the stochastic geometry and the resulting interference further affect the future spectrum access behaviors. This approach provides new insights on the temporal and spatial interactions of spectrum access, which is particularly useful in evaluating the stability of spectrum access protocols or assessing the network robustness to attacks.

\section{Performance Evaluation}
\label{sec_num}

We investigate the dynamics of populations who comply with the rules and make concurrent transmissions under the DoS attack. We are also interested in the averaged SINRs received at PRs and SUs, which reflects the effects of the attack. This section investigates dynamics of access strategies of SUs under the DoS attack. The system parameters are set as $\alpha=4$, $\lambda_{PT}=10^{-5}$, $\eta_{PR}=3$, $\epsilon_{PR}=0.05$, $r_{PT}=15$, $P_{PT}=0.3$, $\lambda_{SU}=10^{-3}$, $\eta_{SU}=3$, $\epsilon_{SU}=0.1$, $r_{SU}=10$, $P_{MU}=P_{SU}=0.1$, $N=10^{-9}$ and $\widehat{\lambda}_{MU}=10^{-7}$.

\begin{figure}[t]
    \centering
    \includegraphics[width=3.5in]{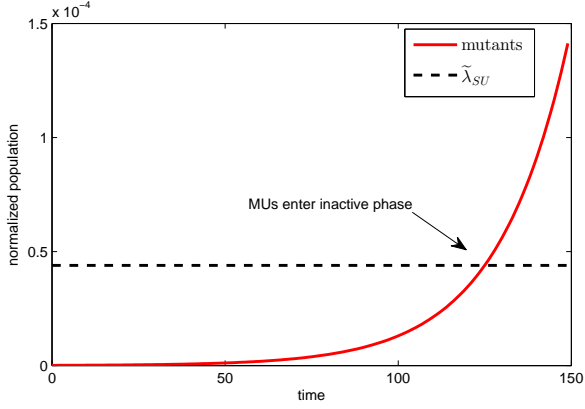}
    \caption{Evolutionary access dynamics when $M=2$, $\delta=10$, $\nu=1$ and $\kappa=0$. Non-mutants represent SUs which stay silent, and mutants represent SUs which transmit for sure. The system parameters are set to be $\alpha=4$, $\lambda_{PT}=10^{-5}$, $\eta_{PR}=3$, $\epsilon_{PR}=0.05$, $r_{PT}=15$, $P_{PT}=0.3$, $\lambda_{SU}=10^{-3}$, $\eta_{SU}=3$, $\epsilon_{SU}=0.1$, $r_{SU}=10$, $P_{SU}=P_{SU}=0.1$, $N=10^{-9}$ and $\hat{\lambda}_{MU}=10^{-7}$. The MUs enter inactive phase when the mutants' population exceed $\widetilde{\lambda}_{SU}$.}
    \label{fig_kappa0}
\end{figure}

\begin{figure}[t]
    \centering
    \includegraphics[width=3.5in]{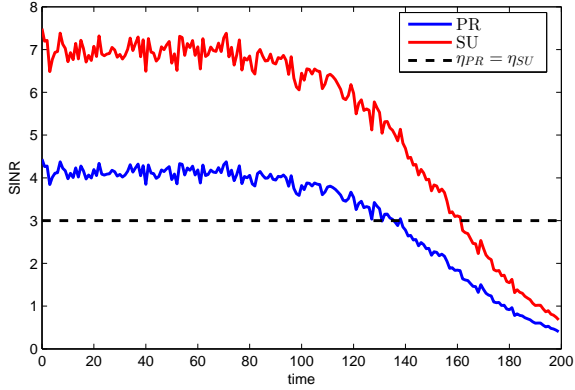}
    \caption{Averaged SINR when $\delta=10$, $\nu=1$ and $\kappa=0$. The system parameters are the same as Figure~\ref{fig_kappa0}.}
    \label{fig_SINR_kappa0}
\end{figure}

Figure~\ref{fig_kappa0} plots the evolution of the fraction of SUs who conduct concurrent transmissions with PUs. Regarding the parameters related to payoffs, we took $(\delta, \nu, \kappa)$ as $(10, 1, 0)$. In the case of only two strategies, non-mutants represent behaved SUs who stick to the rules, and mutants represent misbehaved SUs that transmit for sure. Black line represents the maximum allowable density of SUs when outage constraint of a PR is satisfied (denoted by $\widetilde{\lambda}_{SU}$) and red solid line represents the fraction of SUs simultaneously transmitting with PUs (known as induced SUs). After the population of mutants exceeds the threshold $\widetilde{\lambda}_{SU}$, all primary transmissions fail. Under the same parameter setup, Figure~\ref{fig_SINR_kappa0} plots the averaged SINRs received at a PR and an SU over time. The black dashed line represents that $\eta_{PR} = \eta_{SU} = 3$. Both figures show that if no rewards are supported to SUs for compliance, all SUs will execute concurrent transmissions due to the fact that the original access strategy is not an attractive one for survivability. In this case, the self-enforcement mechanism fails to regulate induced SUs and behaved SUs are extinct. Consequently, significant interference is incurred to PRs and SUs and thus the DoS attack successfully breaks down the CRN.

\begin{figure}[t]
    \centering
    \includegraphics[width=3.5in]{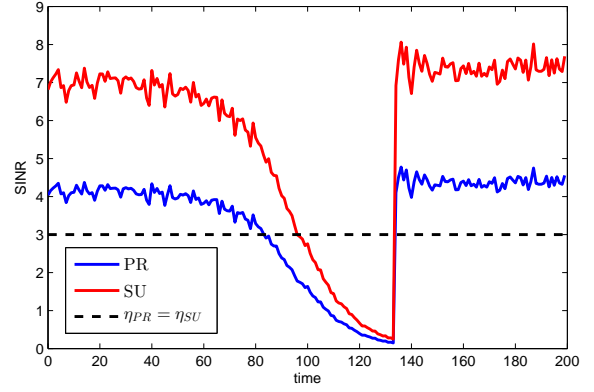}
    \caption{Averaged SINR when $\delta=10$, $\nu=1$ and $\kappa=8$. The positive reward renders SUs stay silent when too many SUs are transmitting simultaneously. The system parameters are the same as Figure~\ref{fig_kappa0}.}
    \label{fig_SINR_kappa8}
\end{figure}

\begin{figure}[t]
	\centering
	\vspace{0.54cm}
	\includegraphics[width=3.5in]{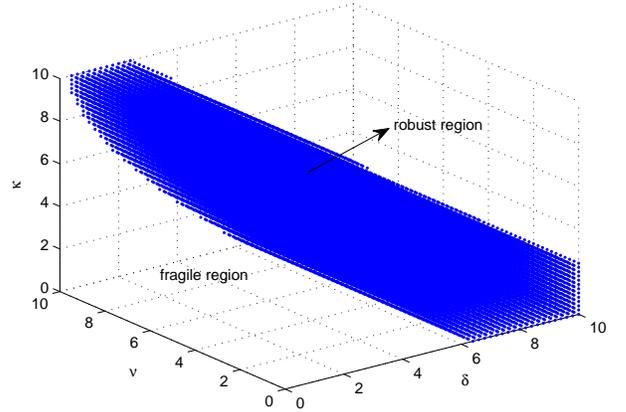}
	\caption{Robust and fragile operating regions under different setups of $(\delta, \mu, \kappa)$.}
	\label{fig_region}
\end{figure}

Figure~\ref{fig_SINR_kappa8} plots the averaged SINRs received at a PR and an SU when rewards are provided (i.e., $\kappa = 8$). This figure shows that at beginning, mutants behave as the dominating species due to the incentives associated with the concurrent transmissions. When too many induced SUs coexist, SUs do not benefit from misbehaving access due to heavy intra-system interference suffered and consequently decide to stay silent to gain reward associated with compliance. By comparing Figures~\ref{fig_kappa0} and~\ref{fig_SINR_kappa8}, we observe that the incentive-base solution is feasible to prevent greedy behavior and thus the proposed attack.

The numerical results of the robust region where the extinction of misbehaved SUs is guaranteed are illustrated in Figure~\ref{fig_region}. With the robust region for ESS identified in Figure~\ref{fig_region}, operator may therefore select appropriate $(\delta, \mu, \kappa)$ so that the CRN could operate in the desirable operating point, in the sense that network breakdown will not occur. On the contrary, the adversary could determine if the proposed attack is feasible according to the predefined $(\delta, \mu, \kappa)$.

\section{Conclusion}
\label{sec_con}
With selfish and greedy nature, SUs may not want to invest effort in complying the sharing rule that no concurrent transmission is allowed, and incur harmful interference to PUs. Inspired by the behaviors of invasive species in an ecosystem coexisted with native species, we propose a novel ecology-based DoS attack where MUs induce original well-behaved SUs to collaboratively transmit by showing them significant incentives to do so. As a result, SUs generate interference to PUs and other SUs, which eventually collapses the entire network. The proposed ecology-based DoS attack is difficult to be detected since, by acting as malfunctioning SUs, MUs are hard to be identified. Using evolutionary access game model, misbehaved SUs are modeled as mutants with distinct access strategy, and the dynamics of access strategies under the attack are analyzed. Numerical results show that the existence of fragile operating region at which SUs are eventually induced to make concurrent transmissions ensures the effectiveness of the attack. The proposed attack demonstrates that when CRs are introduced, the resulting information asymmetry among heterogeneous nodes makes the spectrum sharing mechanism vulnerable and fragile. A robust CRN of cooperation design which is resilient to the proposed inducing attacks is therefore of urgent need in the future.

\end{document}